\documentclass[10pt,twocolumn]{confpaper}

\usepackage{pifont}
\usepackage{algorithm}
\usepackage{algorithmic}
\usepackage{url}
\usepackage{breakurl} 

\newcommand{\system}{{OCDN}}
\def\checkmark{\tikz\fill[scale=0.4](0,.35) -- (.25,0) -- (1,.7) -- (.25,.15) -- cycle;}
\graphicspath{ {./figures/} }

\date{}
\title{\LARGE{\bf OCDN: Oblivious Content Distribution Networks}}
\author{
{Anne Edmundson, Paul Schmitt, Nick Feamster, Jennifer Rexford} \\
{Princeton University}
}

\begin{document}

\maketitle


\begin{sloppypar}

\begin{abstract}  
As publishers increasingly use Content Distribution Networks
(CDNs) to distribute content across geographically diverse networks, 
CDNs themselves are becoming unwitting targets of requests for both
access to user data and content takedown. From copyright infringement to
moderation of online speech, CDNs have found themselves at the forefront of
many recent legal quandaries. At the heart of the tension, however, is the fact
that CDNs have rich information both about the content they are serving and the
users who are requesting that content. 
This paper offers a technical contribution that is relevant to this ongoing tension
with the design of an Oblivious CDN (OCDN); the system is both compatible with the existing
Web ecosystem of publishers and
clients and hides from the CDN both the content it is serving and the users
who are requesting that content. \system{} is compatible with the way that publishers
currently host content on CDNs. Using \system{}, publishers can use multiple CDNs
to publish content; clients retrieve content through a peer-to-peer anonymizing network
of proxies.
Our prototype implementation and
evaluation of \system{} show that the system can obfuscate both content and clients
from the CDN operator while still delivering content with good performance.
\end{abstract}

\ifthenelse{\equal{\onlyAbstract}{no}}{

\section{Introduction}
\label{sec:intro}


As Content Distrubtion Networks (CDNs) host an increasing amount of content
from a diversity of publishers, they are fast becoming targets of requests for
data about their content and who is requesting it, as well as requests for
takedown of material ranging from alleged copyright violations to offensive
content. The shifting legal and political landscape suggests that CDNs may
soon face liability for the content that they host. For example, the European
Union has been considering laws that would remove safe harbor protection on
copyright infringement for online service providers if they do not deploy
tools that can automatically inspect and remove infringing content~\cite{eu-copyright}.  
In the United States, various laws under consideration threaten
aspects of Section 230 of the Communications Decency Act, which protects CDNs
from federal criminal liability for the content that they host. Tussles
surrounding speech, from copyright violations to hate speech, are currently
being addressed in the courts, yet the legal outcomes remain ambiguous and
uncertain, sometimes with courts issuing opposing rulings in different cases.
Regardless of these outcomes, however, CDNs are increasingly in need of {\em technical}
protections against the liability they might face as a result of content that they
(perhaps unwittingly) serve.

Towards this end, we design and implement a system that allows clients to
retrieve web objects from one or more CDNs, while preventing the CDNs from
learning either (1)~the content that is stored on the cache nodes; or (2)~the
content that clients request. We call this system an {\em oblivious
CDN}~(\system{}), because the CDN is oblivious to both the content it is
storing and the content that clients request.

\system{} allows clients to request individual objects with identifiers that
are encrypted with a key that is shared by an open proxy and the origin server
that is pushing content to cache nodes, but is not known to any of the CDN
cache nodes.  To do so, the origin server publishes content obfuscated with a
shared key, which is subsequently shared with a proxy that is responsible for
routing requests for objects corresponding to that URL.  A client forwards a
request for content through a set of peers (\ie, other OCDN clients) in a way
that prevents both other clients  and the CDN from learning the client
identity or requested content.  After traversing one or more client
proxies, an exit proxy transforms the URL that it receives from a client to an
obfuscated identifier using the key that is shared with the origin server
corresponding to the identifier.  Upon receiving that request from the exit
proxy, the CDN returns the object corresponding to the object identifier; that
object is encrypted with a key that is shared between the origin and the
proxy. This approach allows a user to retrieve content from a CDN without
any node in the CDN ever seeing the URL or the corresponding content, or even knowing
the identity of the client that made the original request. Using \system{} requires
only minimal modification to existing clients; clients can also configure aspects
of the system to trade off performance for privacy.

Ensuring that the CDN operator never learns information about either (1)~what
content is being stored on its cache nodes or (2)~which objects individual
clients are requesting is challenging, due to the many possible inference
attacks that a CDN might be able to mount. For example, previous work has
shown that even when web content is encrypted, the retrieval of a collection
of objects of various sizes can yield information about the web page that was
being retrieved~\cite{panchenko2016website, cai2012touching}. Similarly, URLs
can often be inferred from relative popularity in a distribution of web
requests, even when the requests themselves are encrypted. Additionally, the
\system{} design assumes a strong attack model (Section~\ref{sec:threat}),
whereby an adversary can request logs from the CDN, interact with \system{} as
a client, a proxy, or a publisher, and mount coordinated attacks that depend on
multiple such capabilities. Our threat model does not include active attempts
to disrupt the system (\eg, blocking access to parts of the system, mounting
denial of service attacks), but it includes essentially any type of attack
that involves observing traffic and even directly interacting with the system
as a client or a publisher.

The design of \system{} (Section~\ref{sec:design}) under such a strong attack
model entails many unique aspects and features. Because the system allows any
client to join as a proxy, even setting up the infrastructure is challenging.
For example, an attacker could try to join the system as a proxy with the
intent of proxying for specific web content, in an attempt to either disrupt
or surveil those requests. To counter this threat, \system{} uses consistent
hashing to map object identifiers (\ie, URLs) to the proxy responsible for
ultimately routing traffic to the CDN that hosts the object; to ensure that
publishers only communicate keys to the proxies responsible for their content,
each proxy must prove its identity to the respective publisher using a proof
that relies on a self-certifying identifier. 

Requesting and retrieving content, a process that we describe in detail in
Section~\ref{sec:protocol}, is challenging since neither the CDN nor the proxy
must know which client originated a request for a specific piece of content.
The key exchange between an origin server and its respective proxy protects
the confidentiality of both the content and the identifier (\ie, the URL) from
the CDN. To obfuscate the source of the original request, clients construct a
source route to an {\em exit proxy}, but the route can be prepended with
proxies that precede the client who originated the request. To defend against
various inference attacks, as well as to balance load, the \system{} design
allows publishers to use multiple CDNs to distribute the same content,
ensuring that no single CDN has access to information such as the relative
popularity distribution of all objects. To ensure that no single proxy learns
the request pattern for a single object, as well as to balance load, the
design also can also use consistent hashing to assign a set of proxes to a
single object. 

The design of \system{} against a strong adversary is a major
contribution of this work; additionally, we have also implemented \system{}
(Section~\ref{sec:implementation}) and publicly released the source code.
Section~\ref{sec:performance} studies the performance implications of the
tradeoffs between performance and privacy, as well as how \system{} performs
relative to a conventional CDN; Section~\ref{sec:sec} analyzes how \system{}
defends against threats from our adversary.
\ref{sec:discussion} describes various
limitations and possible avenues for future work, Section~\ref{sec:related}
discusses related work, and Section~\ref{sec:conclusion} concludes.

\section{Background}
\label{sec:background}

We now outline how a CDN typically operates, including what information it
has access to by virtue of running a CDN. We also discuss some of the ongoing legal
questions that CDNs currently face.

\subsection{Content Distribution Networks}
CDNs provide content caching as a service to content publishers.  A 
content publisher may wish to use a CDN provider for several reasons:

\begin{itemize}
\item CDNs cache content in geographically distributed locations, which allows for localized 
data centers, faster download speeds, and reduces the load on the content publisher's server.
\item CDNs typically provide usage analytics, which can help a content publisher get a better 
understanding of usage as compared to the publisher's understanding without a CDN.
\item CDNs provide a high capacity infrastructure, and therefore provide higher availability, 
lower network latency, and lower packet loss.  
\item CDNs' data centers have high bandwidth, which allows them to handle and mitigate DDoS attacks better 
than the content publisher's server.
\end{itemize}

CDN providers usually have a large number of edge servers on which content is cached; for example, 
Akamai has more than 216,000 servers in over 120 countries around the world~\cite{akamai_facts}.  
Having more edge servers in more locations increases the probability that a cache is geographically 
close to a client, and could reduce the end-to-end latency, as well as the likelihood of some kinds of 
attacks, such as BGP (Border Gateway Protocol) hijacking.  This is evident when a client requests a web page; the closest 
edge server to the client that contains the content is identified and the content is served from that 
edge server.  Most often, this edge server is geographically closer to the client than the content publisher's 
server, thus increasing the speed in which the client receives the content. If the requested page's content is 
not in one of the CDN's caches, then the request is forwarded to the content publisher's server, the CDN 
caches the response, and returns the content to the client. 

\begin{figure}[t]
\centering
\includegraphics[width=.5\textwidth]{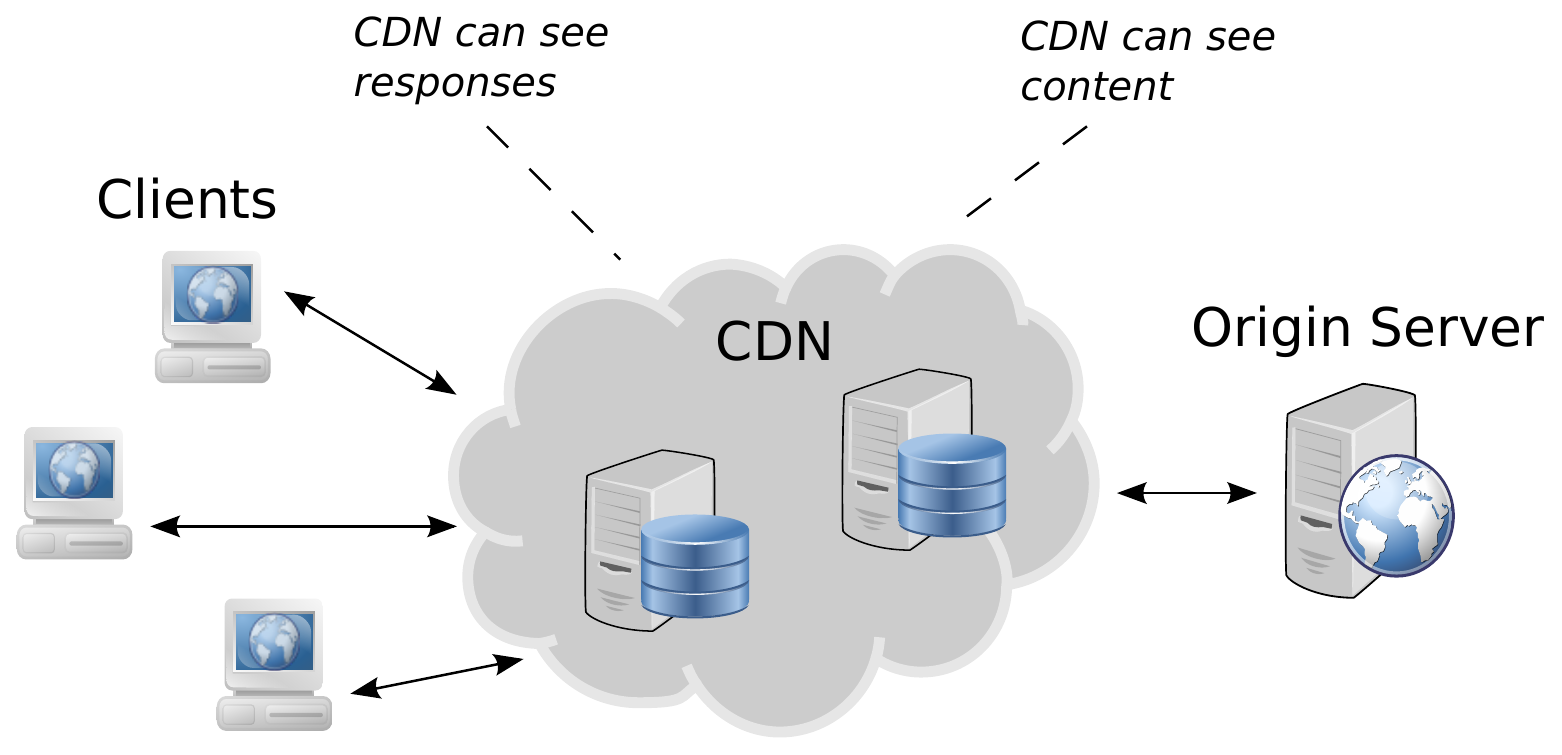}
\caption{The relationships between clients, the CDN, and content publishers in 
CDNs today.}
\label{fig:basic_cdn}
\end{figure}

\subsection{What CDNs Can See}
\label{sec:info}
Because the CDN interacts with both content publishers and clients, as shown in Figure \ref{fig:basic_cdn}, it is in a unique position 
to learn an enormous amount of information.  CDN providers know information about all clients who
access data stored at the CDN, information about all content publishers that cache content at 
CDN edge servers, and information about the content itself.

\paragraph{Content}  CDNs, by nature, have access to all content that they distribute,
as well as 
the corresponding URL.  First, the CDN must use the URL, which is not 
encrypted or hidden, to locate and serve the content. Therefore, it is evident that
the CDN already knows what content is
stored in its caches.  Because CDNs provide analytics to content publishers, they
keep track of cache hit 
rates, and how often content is accessed.  The CDN does not only knows
about the content identifier; it also also 
has access to the plaintext content.  A CDN performs optimizations on the content
to increase performance; 
for example, CDNs minimize CSS, HTML, and JavaScript files, which reduces file sizes by about 20\%.  They can 
also inspect content to conduct HTTPS re-writes; we discuss how \system{} handles these types of optimizations  
in Section \ref{sec:discussion}. In addition, requesting content via HTTPS does not hide any information 
from the CDN; if a client requests a web page over HTTPS, the CDN terminates the TLS connection on behalf of the 
content publisher.  This means that not only does the CDN know the content, the
content identifier, but also it knows 
public and private keys, as well as certificates associated with the content it caches.  


\paragraph{Client information} Clients retrieve content directly from the
CDN's edge servers, which reveals
information about the client's location and what the client is accessing.  
CDNs can also see each client's cross-site browsing patterns: CDNs host content
for
many different publishers, which allows 
them to see content requests for content published by different publishers.  This gives an enormous amount of 
knowledge to CDNs; for example, Akamai caches enough content around the world to see up to 30\% of global Internet 
traffic~\cite{akamai_global_traffic}.  The implications of a CDN having access to
this much information was evident when Cloudflare
went public with the National Security Letters they had received~\cite{cloudflare_nsl};
these National Security Letters
demanded information collected by the CDN and also included a gag order, which prohibits
the CDN from publicly announcing the information request.  

\paragraph{Content publisher information} A CDN must know information
about their customers, the content
publishers; the CDN keeps track of who the content publisher is and 
what the publisher's content is.  The combination of the CDN seeing all content in plaintext and the content's 
linkability with the publisher, gives the CDN even more power.  Additionally, as mentioned previously, the CDN often 
holds the publisher's keys (including the private key!), and the publisher's certificates.  This has led to doubts 
about the integrity of content because a CDN can impersonate the publisher from the client's point of view~\cite{levy2015stickler}.

\subsection{Open Legal Questions}

Various parties are battling in the courts over cases that pertain to user data
requests and intermediary liability.  Large companies
often have large numbers of users, which makes them a target of data requests, for example by a government entity.  Intermediary 
liability would impose criminal liability on an Internet platform (or a CDN) for
the content it provides on behalf of its customers or users.
In the following section, we highlight some of these cases, which all point to a key problem that CDNs face: by knowing all the content 
that they distribute, CDNs may be burdened with the legal responsibility for the
actions of their customers and clients.

\paragraph{User Data Requests}
There are numerous open questions in the legal realm regarding which government can request data stored in different countries, which 
has led to much uncertainty.  A series of recent events have illustrated this uncertainty.  In the struggle over government access to 
user data, cases such as {\it Microsoft vs. United States} (often known as the ``Microsoft Ireland Case'') concerns whether the United 
States Government should have access to data about U.S. citizens stored abroad, given that Microsoft is a U.S. corporation.  

Additionally, there have been user data requests asked of CDNs.  The Cloudflare CDN has been required
to share data with FBI~\cite{cloudflare_nsl}; similarly, leaked NSA documents showed
that the government agency ``collected information `by exploiting inherent 
weaknesses in Facebook's security model' through its use of the popular Akamai content
delivery network''~\cite{facebook_surv}.

\paragraph{Intermediary Liability}
More recently, questions on intermediary liability have been in the spotlight.  For example, many groups, including the Recording Industry 
Association of America (RIAA) and the Motion Picture Association of America (MPAA), have started targeting CDNs with takedown notices for 
content that allegedly infringes on copyright, trademarks, and patent rights; CDNs are a more convenient target of these takedown notices than 
the content provider because oftentimes the content provider is either located in a jurisdiction where it is difficult to enforce the takedown, 
or it is difficult to determine the owner of the content \cite{medium_copyright,eff_copyright}.
Although Section 230 of the Communications Decency Act protects intermediaries,
such as CDNs, from being held
liable for the content they distribute, there have been cases where CDNs are forced
to remove content.  This happened in 2015, as mentioned in Section~\ref{sec:info}, which 
involved the RIAA, Cloudlfare, and Grooveshark \cite{techdirt_copyright}.  
And again in 2017, a district court ruled that Cloudflare is not protected from anti-piracy injuctions by the Digital Millennium Copyright Act (DMCA); the 
RIAA obtained a permanent injunction against a site known as MP3Skull, which contained pirated content, and was distributed by Cloudflare.  The ruling 
did not specify that Cloudflare was enjoined with MP3Skull under the DMCA, but rather that Cloudflare was helping MP3Skull in evading the injunction (under 
Rule 65 of the Federal Rules of Civil Procedure) \cite{stack_copyright}.

The role of a CDN as an intermediary has also come into question in new and currently
pending legislation, including a new German hate speech law and
a bill proposed by the U.S. Senate called Stop Enabling Sex Traffickers Act (SESTA).  In October 2017, Germany passed a new law that imposes large fines, upwards of five million euros,
on social media companies that do not take down illegal, racist, or slanderous comments and posts within 24 hours \cite{nytimes_hatespeech}.  The law 
targets companies such as Facebook, Google, and Twitter, but could also apply to
smaller companies, which could be serviced by CDNs.  In the latter case, it is
an open question whether this new law also applies to CDNs.  In the United States,
the SESTA bill would make Internet platforms liable for their user's illegal comments
and posts \cite{medium_sesta}.  SESTA would hold CDNs liable for the content that
they distribute (despite the CDN not being a party in the content publishing); these
types of laws can naturally lead to overblocking, where an intermediary errs on
the side of caution and censors more content than it needs to.
Such a law may also set a precedent 
for the censoring other types of content that are unpopular but legal. 

\section{Threat Model and Security Goals}
\label{sec:threat}
In this section, we describe our threat model, outline the capabilities of the 
attacker, and introduce the design goals and protections that \system{} provides.

\subsection{Threat Model}
\label{sec:attacker}

Our threat model is a powerful adversary who has a variety of capabilities,
including both surveilling activities and joining the system in various capacities.
We assume that an adversary can gain
access to the CDNs logs, which typically contains client IP addresses and URLs for
each request. Additionally,
the adversary could join \system{} as either a client or any number of clients,
or 
as an arbitrary number of exit proxies.  The adversary could 
also act as an origin server (a content publisher).  We also assume that the adversary
can coordiante several of these actions to learn more information.  For example, the 
adversary could join as a client and an exit proxy, and request access to the CDN's
logs to observe how its own requests are obfuscated.  Additionally, 
the adversary can perform actions, such as generating requests as a client, or creating content 
as a content publisher. The goal of this type of adversary is to learn about the content being stored 
at the CDN and/or learn about which clients are accessing which content.  

The strong adversary that we consider has seen some precedent in practice:
for example, governments have demanded access to CDNs' 
data~\cite{cloudflare_nsl}.  Although one possible adversary is a government requesting
logs from the CDN, 
the government could also be colluding with a CDN; the CDN operator might even be
an adversary.

Our design does not defend against an attacker who attempts to actively disrupt
or block access to the system, such as by actively
modifying content, disrupting communications (\eg, through denial of service), or
blocking access, content, or requests. Prior work on securing
CDNs has introduced methods to handle an actively malicious adversary by preserving the integrity of content 
stored on CDN cache nodes~\cite{levy2015stickler}.  We do not address an adversary that tampers, modifies, or 
deletes any data, content, or requests.  

\subsection{Security and Privacy Goals for \system{}}
\label{sec:goals}

To defend against the adversary described in Section~\ref{sec:attacker}, we highlight
the design goals for \system{}. 
Each stakeholder---in this case the content publisher, the CDN, and the client---has
different risks, and therefore should have different protections.  All three stakeholders 
can be protected by preventing CDNs from learning information, decoupling content distribution from trust, and 
maintaining the performance benefits of a CDN while reducing the probability of attacks.  
One strength of \system{} is that it protects the origin server, the CDN itself,
and the client, whereas existing systems, such as Tor, only protect the client.



\paragraph{Prevent the CDN from knowing the content it is caching} First and foremost, the CDN 
should not have access to the information outlined 
in Section~\ref{sec:background}.  By limiting the information that the CDN knows,
\system{} limits 
the amount of information that an adversary can learn or request.  \system{} should hide 
the content as well as the URL associated with the content.  If the CDN 
does not know what content it is caching, then the CDN will not be able to supply an adversary 
with the requested data and it will have a strong argument as to why it cannot be held 
liable for its customers' content.

\paragraph{Prevent the CDN from knowing the identity of users accessing content} CDNs can currently see clients' 
browsing patterns. \system{} should provide privacy protections by hiding which client is accessing 
which content at the CDN.  In addition, it should hide cross-site browsing patterns,
which a CDN 
is unique in having access to.  Some CDNs block legitimate Tor users because they are 
trying to protect cached content from attacks, such as comment spam, vulnerability scanning, 
ad click fraud, content scraping, and login scanning~\cite{ars_tor}; for example,
Akamai blocks Tor users~\cite{khattak2016you}. As a positive side effect, \system{}
prevents
privacy-conscious Tor users from being blocked by CDNs.  Finally, some CDNs, due
to their ability
to view cross site browsing patterns, could de-anonymize Tor users~\cite{cloudflare_tor}, but \system{} would 
prevent a CDN from compromising the anonymity of clients.

\subsection{Performance Considerations}
As one of the primary functions of a CDN is to make accessing content faster and more 
reliable, \system{} should consider performance in design decisions.  The performance of \system{} will 
be worse than that of traditional CDNs because it is performing more operations on content, but \system{} 
is offering confidentiality, whereas traditional CDNs are not.  \system{} should scale linearly in terms of load 
and storage requirements on exit proxies; additionally, it should be able to 
scale with the number of clients using the system, as well as with the growing number of web pages on the internet.

\section{Design}
\label{sec:design}

\system{} that provides oblivious content distribution and  
comprises the following components: clients, exit proxies, CDNs, and origin 
servers.  {\em Clients} are the Internet users who use the system to access content
stored on CDN cache nodes; {\em exit proxies} are proxies that obfuscate the requests
and responses retrieved from the CDNs; and the {\em origin servers} are the content
publishers who are customers of the CDNs.  Figure \ref{fig:ocd_overview} shows how
these components interact in the system.  This section describes the decisions 
made in the design of \system{}, and what functionality each decision provides.  
We separate design decisions into two parts: 1)~setup decisions and 2)~request/response

decisions.  We also highlight some additional options that the design of 
\system{} allows.

\begin{figure}[t!]
\centering
\includegraphics[width=.5\textwidth]{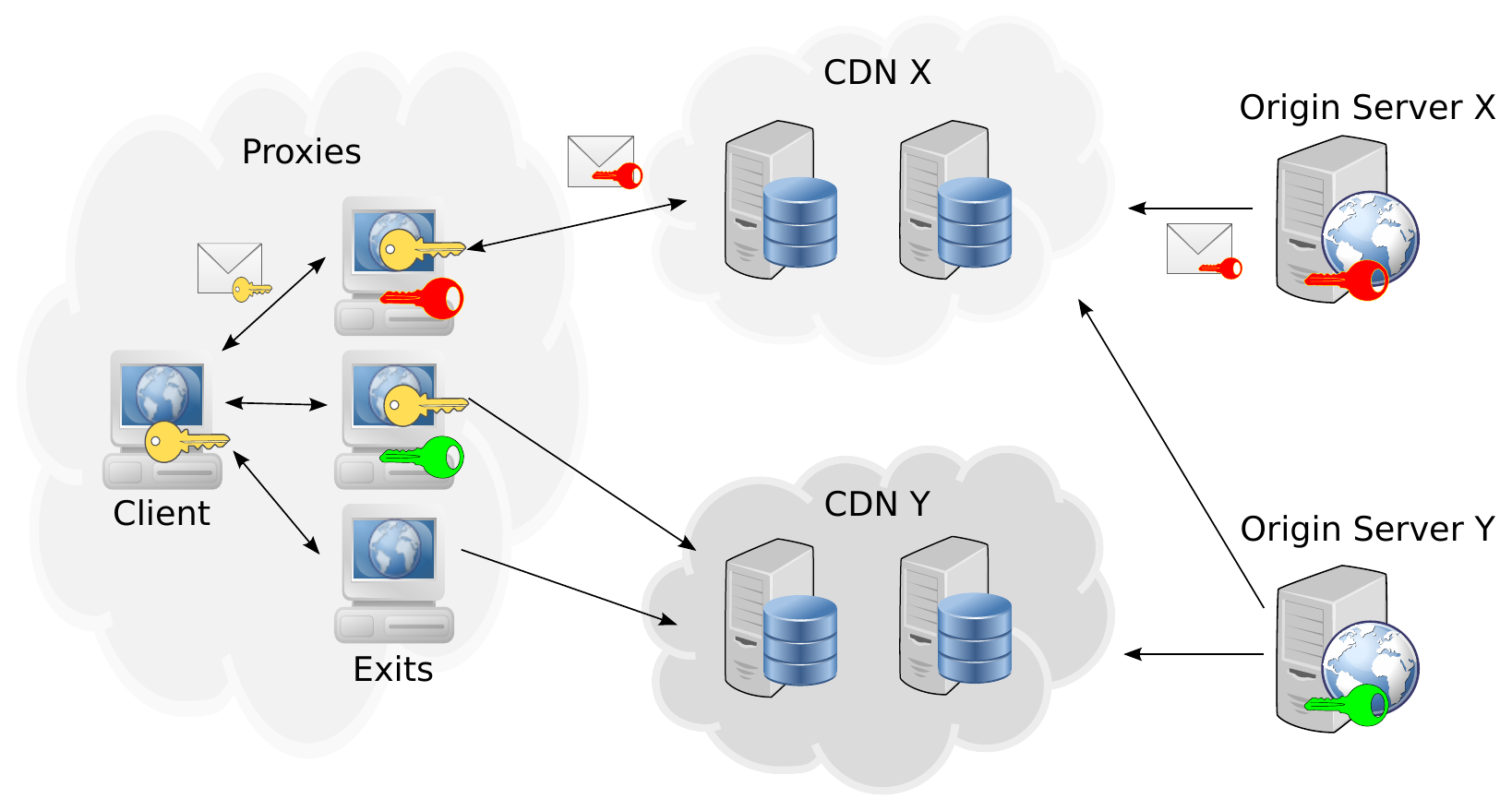}
\caption{The relationships between clients, exit proxies, CDNs, and origin servers in 
\system{}.}
\label{fig:ocd_overview}
\end{figure}

\subsection{\system{} Setup}
We start by discussing how the system components communicate and authenticate 
one another; Table~\ref{tab:setup} summarizes these decisions. 
We introduce shared keys between origin servers and exit proxies, how these keys
are 
stored, how the exit proxies authenticate themselves to origin servers, and how these 
keys are distributed.

\begin{table}[t!]
\footnotesize
\centering
\begin{tabular}{ l  p{2in} } 
 \multicolumn{1}{c}{\bf Design Decision} & \multicolumn{1}{c}{\bf Function} \\
\hline \hline
 Shared Keys & {Hides content on cache nodes from CDN.} \\
 Consistent Hashing & {Load balance requests across proxies; ensure no proxy can
 control a given URL.} \\
 Self-Certifying Identifiers & {Authenticates exit proxies to origin servers.} \\
 DNS for Key Sharing & {Allows origin server to share shared keys with exit
 proxies.} \\ \hline
\end{tabular}
\caption{Design decisions associated with \system{} setup.}
\label{tab:setup}
\end{table}

\paragraph{Shared Keys} 
To prevent an adversary from learning information, the CDN must not know anything
about the
content that it is caching.  Therefore, the content {\it and} the associated URL
must be obfuscated
before the CDN sees them.  The content can be obfuscated by encrypting it with a
key that is not
known to the CDN.  Because this must be done prior to any caching, the content publisher must 
generate a shared key $k$ to encrypt the content with. Encrypting the content alone does not 
hide much from the CDN; the content identifier, or URL, must also be obfuscated, otherwise the 
CDN can still reveal information about which clients accessed which URLs (which is indicative 
of the content).  In obfuscating the URL, the result should be fixed and relatively
small; 
these requirements reduce storage requirements and prevent the adversary from guessing
the
URL based on the length of the obfuscated URL.  Unfortunately, using a simple hash allows an 
attacker to guess the content identifier by hashing guesses and comparing with 
the hashes stored in the CDNs caches.  Therefore, the content publisher incorporates the use 
of the shared key $k$ into the hash of the URL by using a hash-based message authentication code 
(HMAC).  Additionally, if the domain supports HTTPS requests, then the content publisher must 
also encrypt the associated certificate with the same key $k$.

The encrypted content and corresponding HMAC are sent to the CDN\footnote{Most CDNs
allow the publisher to
decide on a push or pull model, but tje \system{} is compatible with either approach.}
and stored in
its caches.  The content publisher then shares the key $k$ with an exit proxy. 
This key allows the 
exit proxy to request encrypted content on behalf of clients by computing the HMAC on the URL.  

\paragraph{Consistent Hashing}
Each exit proxy stores a mapping of URLs to their associated shared key $k$; for example, if 
an origin server has shared key $k$ and publishes a web page {\tt www.foo.com}, then an exit 
proxy will store the mapping of {\tt www.foo.com} to $k$.  This results in the set of exit proxies 
forming a distributed hash table where the key is the URL ({\tt www.foo.com}) and the value is the 
shared key ($k$).  To assign (key,value) pairs to exit proxies, \system{} uses consistent 
hashing~\cite{karger1997consistent,lewin1998consistent}.  Consistent hashing uses a hash function $H(.)$
to generate identifiers for both exit proxies and for URLs; the identifiers are $H(exit\_ID)$ and $H(URL)$. 
We discuss what $exit\_ID$ is in the next section on Self-Certifying Identifiers.  After the hashes are 
computed, then they are mapped to a point on an identifier circle (modulo 2$^{m}$, where $m$ is the length of 
identifier in bits); each URL ($H(URL)$) on the circle is assigned to the first exit proxy ($H(exit\_ID)$) that 
is equal to or follows $H(URL)$ on the circle.  This hashing method is used in \system{} because it provides: 
1)~an evenly distributed mapping of URLs to shared keys among the exit proxies,
2)~a way to prevent an exit 
proxy from choosing which URL it wishes to be responsible for, and 3) a relatively small amount 
of (key,values) to be moved when a new exit proxy is established (or removed).  

\paragraph{Self-Certifying Identifiers}
As mentioned in the previous paragraph, consistent hashing makes use of identifiers for both the URLs and 
the exit proxies.  While the identifiers for URLs are straightforward ($H(URL)$), the identifiers for exit 
proxies must provide more information; an exit proxy identifier must be able to prove to an origin server that 
it is the exit proxy that is responsible for the associated URL.  If this validation was not part of \system{}, 
then any (potentially malicious) exit proxy could request the shared key $k$ from any or all origin servers.  To 
prevent a malicious exit proxy from learning any shared key $k$, it must be identified by a self-certifying 
identifer.  This technique was first introduced in a self-certifying file system~\cite{mazieres2000self}; it allows
for other entities (such as origin servers) to certify the exit proxy solely based on its identifier.  The format 
of this identifier ($exit\_ID$) is {\tt IP:hostID}, where {\tt IP} is the exit proxy's IP address and {\tt hostID} 
is a hash of the exit proxy's public key.  When an exit proxy is requesting the shared key $k$ from an origin server, 
it sends its identifier and its public key to the origin server.  The origin server
can then hash the exit proxy's 
public key and verify it against the {\tt hostID}; this action serves as a proof
of the exit proxy's position in the consistent hashing
circle, and thus prevents a proxy from lying about where it lies on the ring (and subsequently lying about which 
URL's shared key it is responsible for).

\paragraph{DNS for Key Sharing}
We have discussed how shared keys are generated, used, and stored, and here we describe how they are shared.  As previously 
stated, the origin servers generate shared keys and must share them with the (correct) exit proxies.  \system{} uses DNS
to do so.  To retrieve a shared key $k$, an exit proxy sends a DNS query to the origin server's authoritative DNS, and 
it includes its identifier, $exit\_ID$, and its public key in the {\tt Additional Info} section of the query.  The 
authoritative DNS for the origin server validates the exit proxy by hashing the public key and comparing it to the 
second part of $exit\_ID$, and verifying that the exit proxy is responsible for its URL based on the consistent 
hashing circle.  If the verification is successful, then the authoritative DNS sends the shared key $k$ encrypted 
under the exit proxy's public key, \{$k$\}$_{PK_{exit}}$ in the SRV record of the DNS response.  The exit proxy 
extracts $k$ by decrypting with its private key, and stores it in its hash table.

\subsection{Requests \& Responses}
We make additional design choices that concern the requests that clients initiate
and
the responses they receive.  Table~\ref{tab:request_response} highlights these decisions;
we 
introduce session keys, how requests are routed from clients to exit proxies, and how responses 
are routed from exit proxies back to the original client.

\begin{table}[t!]
\footnotesize
\centering
\begin{tabular}{ l  p{1.95in} } 
 \multicolumn{1}{c}{\bf Design Decision} & \multicolumn{1}{c}{\bf Function} \\
\hline \hline
Spoofed Source Routes & {Hides origin of client request from other
 clients, exit proxies, and CDN.} \\
 Session Keys & {Hides URL and response from other clients.} \\
 Multicast Response & {Allows CDN to return content directly to client without knowing
 the client that requested the content.} \\
 \hline
\end{tabular}
\caption{The design decisions associated with content requests and responses, and what these 
decisions provide.}
\label{tab:request_response}
\end{table}

\paragraph{Potentially Spoofed Source Routes}
As previosuly described, exit proxies query the CDN on behalf of clients, but the
exit proxy
should not be able to learn which client sent which request.  This obfuscation is
accomplished by routing requests through
a series of other clients.  In \system{}, each client is running a proxy and is
also a peer in this system; this 
peer-to-peer system of clients borrows the 
protocols used for clients joining, leaving, and learning about other clients from
the vast literature on peer-to-peer systems. A client routes a request through
her peers by using source routing; when the client generates a request, it also
generates a source route, which includes
the addresses of a set of her peers.  The last hop in the source route is the exit proxy that is responsible for the 
shared key $k$ associated with the URL in her request.  The client determines the correct exit proxy by looking this 
up in a local mapping (which is retrieved from a central system that keeps the mapping of URLs to exit proxies).  
It appends this source route to its request and forwards it to the next peer
in the route.  When a peer receives
a request, she simply forwards it on to the next peer; this continues until the last hop in the source route, which 
is an exit proxy. 

Although it might initially appear as if it is easy to identify the client
that initiatived a request as the first hop in the source route, \system{}
allows each client to spoof source routes; specifically, a client can prepend
other peers in the route before it initiates a request.  For example, a client
with identity {\it C} could generate a route to exit proxy {\it E} that looks
like $C \rightarrow G \rightarrow F \rightarrow E$ and can further obfuscate
the source of the route by prepending additional clients to the beginning of
the route as follows: $D \rightarrow A \rightarrow C \rightarrow G \rightarrow
F \rightarrow E$.

\noindent Neither {\it G}, {\it F}, nor {\it E} know who the original requestor was; from {\it E}'s point of 
view, the original requestor could have been {\it D}, {\it A}, {\it C}, {\it G},
or {\it F}.  Using a sequence of 
peers, or even just knowing that a client {\it can} use a series of peers, hides
the identity of the client 
from other clients, exit proxies, and the CDN. 

\paragraph{Session Keys}
In addition to shared keys between origin servers and exit proxies, \system{} uses session keys shared 
between clients and exit proxies.  Session keys provide confidentiality of the requested URL and the 
response.  When the client generates a request, it generates a session key $skey$.
and encrypts 
the URL in her request with this key, which provides \{URL\}$_{skey}$.  The client
must also share this session key
with the exit proxy, so that the exit proxy can learn the plaintext URL and subsequently compute the HMAC to 
query the CDN.  The client encrypts the session key with the exit proxy's public key, result in \{skey\}$_{PK_{exit}}$, 
and appends this value as an additional header on the request.  Because her request
could be forwarded through 
a set of client peers, this hides the URL of the request from other clients.

When an exit proxy receives a request from a client, it first extracts the session
key $skey$ by decrypting it with 
his private key, and then he decrypts the URL with the session key.  This operation
yields the original plaintext
URL. Using the shared key $k$ from the origin server, it can then compute
HMAC$_k$(URL) and forward the request 
to the CDN.  Upon receiving a response from the CDN, the exit proxy then decrypts
the content with the shared key $k$, and
encrypts the content with the session key $skey$ before sending it to the client.
When it receives the encrypted response, 
the client can then decrypt it using $skey$.

\paragraph{Multicast Responses}
Using session keys allows for a performance optimization in sending responses back to clients.  Instead of sending 
the encrypted response from the exit proxy back to the client via the set of peers used in the source route, the exit 
proxy can send it in a multicast manner to all clients that were on the source route.  The only client that knows $skey$ 
is the true client that originated the request, therefore none of the other clients can interpret the response, and it reduces the 
latency for sending the response to the client.  

\subsection{Additional Options}
Up to this point, we have discussed how \system{} is designed in the general case.  Here we describe some 
additional options that \system{}'s design can include.

\paragraph{Multiple CDNs}
While describing the design decisions that went into \system{}, we referred to a single CDN for 
simplicity.  In reality, \system{} allows for many CDNs to participate;
distributing content across
multiple CDNs could provide additional privacy. Origin servers can also take advantage
of multiple CDNs.

\paragraph{Encoding URLs}
As described earlier, each URL is obfuscated by using a HMAC and then stored on the CDN.  An adversary 
could potentially correlate a URL's popularity with its access patterns.  To prevent this, \system{} allows 
origin servers to generate multiple different encodings of its URLs, such that HMAC$_k$(enc$_1$(URL)) $\neq$ 
HMAC$_k$(enc$_2$(URL)).  Each origin server could produce $n$ different encodings of popular URLs, such that 
the popularity distribution seen by an adversary is a uniform distribution of URL requests across all URLs.  

\paragraph{DHT Replicas}
Each exit proxy's hash table can be replicated by another (or many other) exit proxies.  This 
would provide less load per exit proxy, as well as redundancy in case of failures.  Additionally, 
the CDN can cache the content associated with a given URL at more than one cache
node;
if only one exit proxy is responsible for a given URL's content, then it would likely only be cached at 
cache node closest to the exit proxy.  Having multiple exit proxies responsible for a URL's content 
helps decrease the load on the proxies while maintaining some of the performance benefits of a CDN.

\paragraph{Partial Content}
Different origin servers have different needs, and each origin server might 
have different needs for different content.  The design of \system{} allows origin servers
 to publish some of their content on \system{} and some on other CDNs.  
This is useful in a case where some content is more sensitive, while other content needs 
better performance.

\paragraph{Pre-Fetch DNS Responses} 
One way to increase the performance of \system{} is to pre-fetch DNS responses at 
the exit proxies.  This would allow the exit proxy to serve each client request faster 
because it would not have to send as many DNS requests.  Pre-fetching DNS responses would 
not take up a large amount of space, but it also would not be a complete set of all DNS 
responses.  Additionally, if the content is moved between cache nodes at the CDN, then DNS 
response must also change; therefore, the pre-fetched DNS responses should have a lifetime 
that is shorter than the lifetime of the content on a cache node.

\paragraph{Privacy vs. Performance Tradeoffs}
There are two different modes that \system{} can operate in, where one provides better 
performance, and the other provides better privacy.  In the first mode, the client can 
choose to send a request directly to the exit proxy.  

In this case, the exit proxy might be able to discover the identity of the client,
but the CDN would still not be able to map a request to the client that made the
request.
Alternatively, the client can forward a request
through a set of peers before it reaches the exit proxy.  In this case, the client
can 
prepend other clients' identifiers (as previously described) to make it appear as
though the request came from
a different client.  This action further obscures the relationship between the client
and the request.  As another option, the client could {\it only} prepend
other clients'identifiers but simply forward
the request {\em directly} to the exit proxy; this action provides the same performance
benefit as the first
mode, but still offers some additional privacy benefits.  Although the last option
would appear to strike the optimal balance between privacy and performance, it cannot
be
the only option because the exit proxy would always know that the true client is the previous hop 
in the source route.  These modes of operation provide clients with different ways to use 
the system both based on their privacy preferences and the type of content they are requesting.

\begin{figure}[t!]
\centering
\includegraphics[width=.45\textwidth]{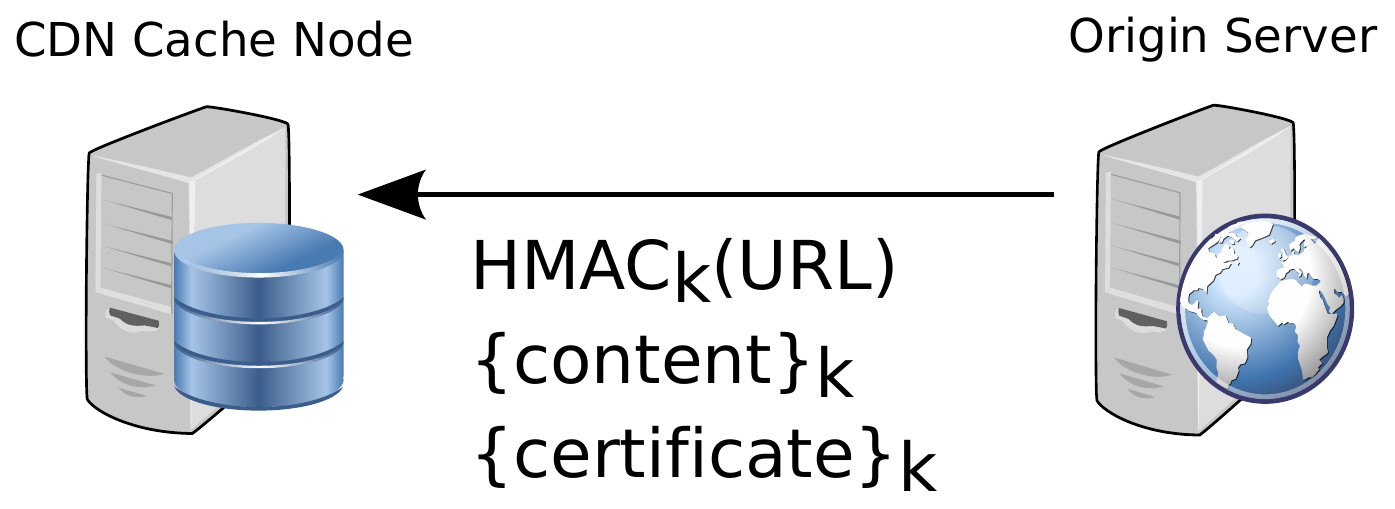}
\caption{How content is published in \system{}.  $k$ is shared between the 
origin server and the corresponding exit proxy; the CDN has no knowledge of $k$.}
\label{fig:publishing}
\end{figure}

\section{\system{} Protocol}
\label{sec:protocol}
Based on the design decisions discussed in the previous section, we specify the 
steps taken to publish and retrieve content in \system{}.

\subsection{Publishing Content}
\label{sec:publish_protocol}
In order to publish content such that the CDN never sees the content, the publisher 
must first obfuscate her content, as described in Section \ref{sec:design}. 
Figure \ref{fig:publishing} shows the steps taken to publish content.


The most important step in content publishing is obfuscating the data.  We assume that the origin 
server already has a public and private key pair, as well as a certificate.  To obfuscate the data 
the origin server will need to generate a shared key $k$. 

Once the key is established, the origin server must first pad the content to the same size for some 
range of original content sizes (i.e., if content is between length x and y, then pad it to length 
z).  The range of content sizes should be small, such that this causes negligible padding overhead, but 
reduces the probability of identifying the content based on the content length.  This content padding 
is done to hide the original content's length, as it may be identifiable simply by its length.  After 
content is padded, then the content is divided into fixed size blocks and padded to 
some standard length.  Then each block is encrypted using the shared key $k$, 
resulting in a set of encrypted blocks. Because the CDN does not have access to the shared key, 
it cannot see what content it is caching.  

Now that the content is obfuscated, the origin server must also obfuscate the content's identifier.  To do so, 
she computes the HMAC of the URL using the shared key $k$.

Once the identifier and the content are obfuscated with $k$, they can be pushed to the CDN, or optionally to multiple 
CDNs.  Recently, services have cropped up to allow and help facilitate the use of multiple CDNs for the same content; an 
origin server could use multiple CDNs' services.  This mechanism could be used in \system{} to increase reliability, 
performance, and availability; an origin server can use a service, such as Cedexis~\cite{cedexis}, to load balance between 
CDNs.  We discuss the use of multiple CDNs more in Section \ref{sec:partial} on \system{} in 
partial deployment.

As the exit proxies use consistent hashing to divide keys among proxies while balancing load, the origin server
determines which exit proxy is correct (based on the consistent hashing circle).  The origin server then encrypts the 
shared key $k$ with the correct exit proxy's public key $PK_{exit}$.  
Figure \ref{fig:keys} shows the steps for retrieveing a shared key.  First, the
exit proxy sends a DNS request to the origin server's authoritative DNS server, including its self-certifying
identifier and its public key (these are both included in the {\tt Additional Info} section of the DNS message).  The origin server hashes the exit 
proxy's public key and verifies it against its self-certifying identifier; this acts as a proof of the 
exit proxy's position in the consistent hashing circle.  If the origin is able to certify the exit proxy, then it will send the DNS response with 
\{$k$\}$_{PK_{proxy}}$ in the SRV record. The exit proxy will receive the encrypted shared key, which it can decrypt with it's private key.

\paragraph{Updating Content}
For an origin server to update content, she must follow similar steps as described in publishing content.  
Once she has updated the content on her origin server, she must obfuscate it using the same steps: 1) pad the 
original content length, 2) divide the content into fixed size blocks, and 3) encrypt the content blocks 
with the shared key $k$.  Because she is updating the content (as opposed to creating new content), the 
obfuscated identifier will remain the same.  The origin server signs the updated obfuscated content with 
her private key, such that the CDN can verify it was true origin server that sent the update.

\paragraph{Updating Keys}
An origin server must be able to update keys in case of compromise.  To minimize the amount of time a key is compromised for, the 
origin server specifies an expiration date and time for the key when it is originally generated.  The origin server 
periodically checks if the key is valid or not based on the expiration timestamp.  If the key is still valid, the origin server 
continues to use it.  Otherwise, the origin server generates a new key $k_{new}$, computes HMAC$_{k_{new}}$(URL), and 
encrypts the content (and possibly certificate) with $k_{new}$.  The content publisher then follows the same steps as in Updating 
Content to push the content to the CDN, and it publishes $k_{new}$ encrypted with the exit proxy's public key in it's DNS SRV record.

The corresponding exit proxy must also be able to fetch this new key $k_{new}$ and replace the expired key with it.  When the exit proxy 
sees an incoming request for a URL that uses key $k$, it first checks $k$'s timestamp.  If valid, then it continues as normal.  Otherwise, 
it sends a DNS request to the publisher's authoritative DNS, and extracts \{$k_
{new}$\}$_{PK_{proxy}}$ from the DNS response.  The exit 
proxy then decrypts it to obtain $k_{new}$, updates its version of the key, and
proceeds as normal.

\subsection{Retrieving Content}
\label{sec:retrieve}
The steps for a client to retrieve a web page that has been cached by \system{} are shown in Figure \ref{fig:retrieving}, where the client 
forwards a request directly to an exit proxy; Figure \ref{fig:retrieving2} shows
the steps to retrieve content when the client forwards its
request through two peers. We assume the client has already joined the system, which is described in more detail in Section \ref{sec:join}; at this 
stage, the client has knowledge of a subset of its peers (other \system{} clients) and a mapping of exit proxies and which URLs they hold 
keys for.  The client generates a request for a specific URL, and looks up which exit proxy holds the key for that URL in its local mapping.  Next, 
the client selects a source route; this source route allows the client to specify which mode of \system{} they would like to use: 1) no additional source 
route, which has better performance, or 2) a source route, which has better privacy.  If the client decides to use the privacy-preserving mode, 
then she generates a source route, which includes some of its peers, and could potentially
include a false originator (as described in Section \ref{sec:design}).
Before sending the request, the client generates a session key $k_{session}$ and encrypts it with the exit proxy's public key.  The client appends both the source route and \{$k_{session}$\}$_{PK_{proxy}}$ to the request and encrypts the URL with $k_{session}$ such that no other clients on the path can learn what the requested URL is.  The client then sends it onto the next proxy in the source route, which could be either another client proxy or the exit proxy.  The request is forwarded 
through all subsequent hops in the source route until it reaches the exit proxy.  The exit proxy decrypts \{$k_{session}$\}$_{PK_{proxy}}$ with its private key and stores 
the source route locally; it then decrypts the URL with $k_{session}$. 

The exit proxy then resolves the domain using it's local resolver, which will redirect it to the CDN's DNS resolver. In order for the exit proxy to 
generate the obfuscated identifier to query the cache node for the correct content, 
it must have the shared key $k$ that the origin server generated and obfuscated the content and identifier 
with.  The steps an exit proxy takes to retrieve the shared key were outlined in Section \ref{sec:publish_protocol} and are shown in Figure \ref{fig:keys}.

Now that the proxy has obtained the shared key $k$ from the origin server, it can generate the obfuscated content identifier based 
on the request the client sent.  It computes the HMAC of the URL with the shared key.  The proxy then 
sends the (obfuscated) request to the edge server, where the CDN locates the content associated with the identifier.  The CDN returns 
the associated obfuscated content, which we recall is the fixed-size blocks encrypted with the same shared key that the identifier was 
obfuscated with.  The proxy can decrypt the content blocks with the shared key from the origin server, assemble the blocks, and strip any 
added padding, to reconstruct the original content.

Lastly, the exit proxy must send the response back to the correct client without knowing who the client is.  First, the exit proxy fetches the session key $k_{session}$ 
that it stored for the corresponding incoming request, and it uses this key to encrypt the response.  Then, it looks up the source route it stored for the corresponding request 
and uses a multicast technique to send to the encrypted response to all clients on the source route.  At this point, the exit proxy can delete the source route and session key entries 
for this request/response.  Only the original (true) client has $k_{session}$, so only the original (true) client can decrypt the response.  All other clients will discard 
the encrypted response because they cannot decrypt it.  

\subsection{Clients Joining \& Leaving}
\label{sec:join}

When a client joins \system{}, she will download \system{} client software.  This includes information about exit proxy mappings to URLs for which they hold a key, 
software for modifying requests with session keys and source routes, and software for running a proxy.  Clients will learn about other clients in the system via 
a gossip protocol.  We do not detail this as gossip protocols have been studied extensively in the past.  Similarly, when a client leave the system, this 
information is propogated to its peers using a gossip protocol.

\subsection{Partial Deployment}
\label{sec:partial}
\system{} should be partially deployable, in the sense that if only some origin servers participate or only some CDNs participate, then 
the system should still offer some protections.  We outline two different partial
deployment possibilities below.

\begin{figure}[t!]
\centering
\includegraphics[width=.4\textwidth]{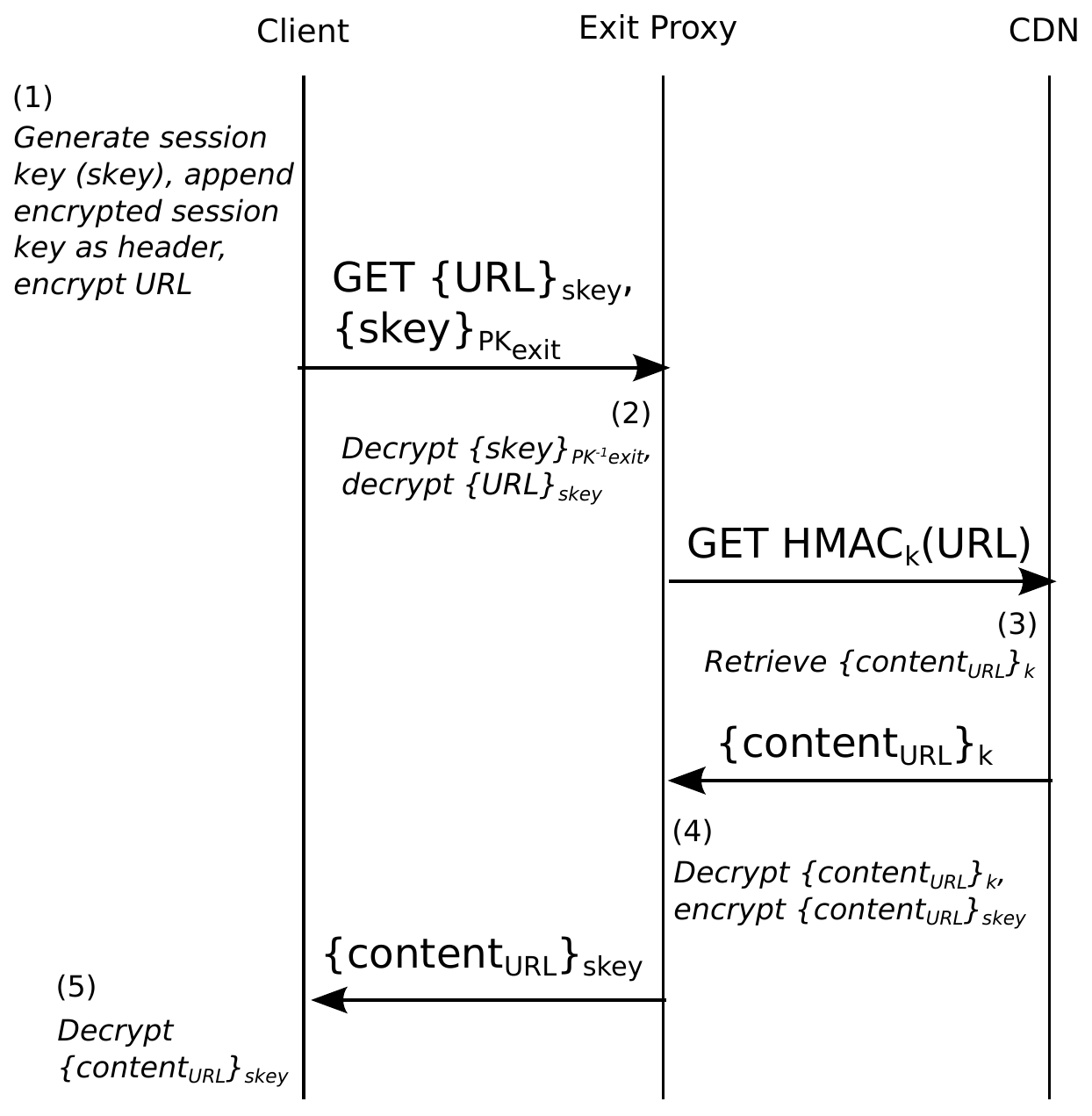}
\caption{Steps for retrieving content in \system{} when a client is prioritizing 
performance and goes directly to an exit proxy.}
\label{fig:retrieving}
\end{figure}

\begin{figure}[t!]
\centering
\includegraphics[width=.5\textwidth]{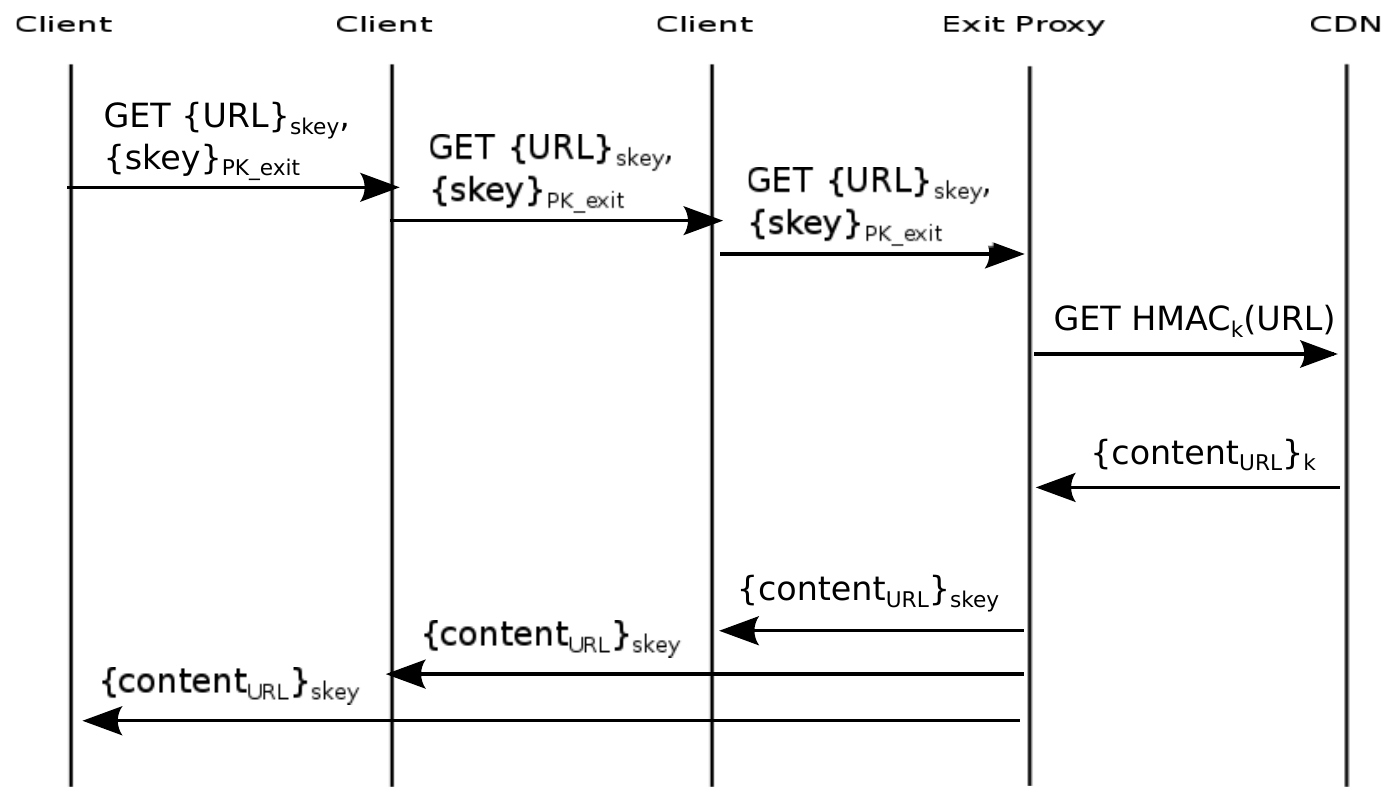}
\caption{Steps for retrieving content in \system{} when a client is prioritizing 
privacy and proxies a request through two other clients before reaching the 
exit proxy.  This figure shows that the request is sent sequentially through peers, 
and the response is sent in a multicast manner back to the clients.}
\label{fig:retrieving2}
\end{figure}

\paragraph{Deployment with Origin Servers' Full Participation}
One option for deploying \system{} is to ensure there is some set $S$ of origin servers that participate fully in the 
system.  These publishers obfuscate their content, identifiers, and certificates, and most importantly, only have 
obfuscated data stored on the CDNs cache nodes.  $S$ must be greater than one, otherwise the CDN can infer 
that a client accessing this obfuscated content is actually accessing content that can be identified.  This partial deployment plan 
 protects the privacy of the clients accessing the content created by the set of origin servers $S$.  It does not 
protect the clients' privacy as completely as full participation of all origin servers in \system{} because the CDN can 
still view cross site browsing patterns among the origin servers that are not participating. It is important to note though, that 
because the clients are behind proxies, the CDN cannot individually identify users.  The CDN can attribute requests to exit proxies, but 
not to clients.  

\paragraph{Deployment with Origin Servers' Partial Participation} 
Some origin servers may prioritize performance 
and availability.  Therefore, they should have the option to gradually move towards full participation by pushing 
both encrypted and plaintext content to the CDN.  In this partial deployment plan, we see some set of origin servers 
fully participating with only encrypted content, some other set of origin servers partially participating with both 
encrypted and plaintext content, and some last set of publishers that are not participating.  Unfortunately, if 
a publisher has the same content that is both encrypted and plaintext content at a cache node, then an adversary can correlate the access 
patterns on encrypted and plaintext content for the origin server.  In order to prevent this identification of the 
content, \system{} can use encoded URLs (described in Section \ref{sec:design}), which obfuscates the access patterns for 
a given piece of content; this holds true if an origin server chooses to distribute its content in an encrypted manner using \system{}
 and in plaintext form on a different CDN.  In this case, the origin server can still encode its URLs in multiple ways to 
prevent correlating access patterns between the encrypted and plaintext content.  Therefore, this deployment option allows for 
differing levels of participation in the system, while still preserving the protections provided by \system{}.  

\begin{figure}[t]
\centering
\includegraphics[width=.45\textwidth]{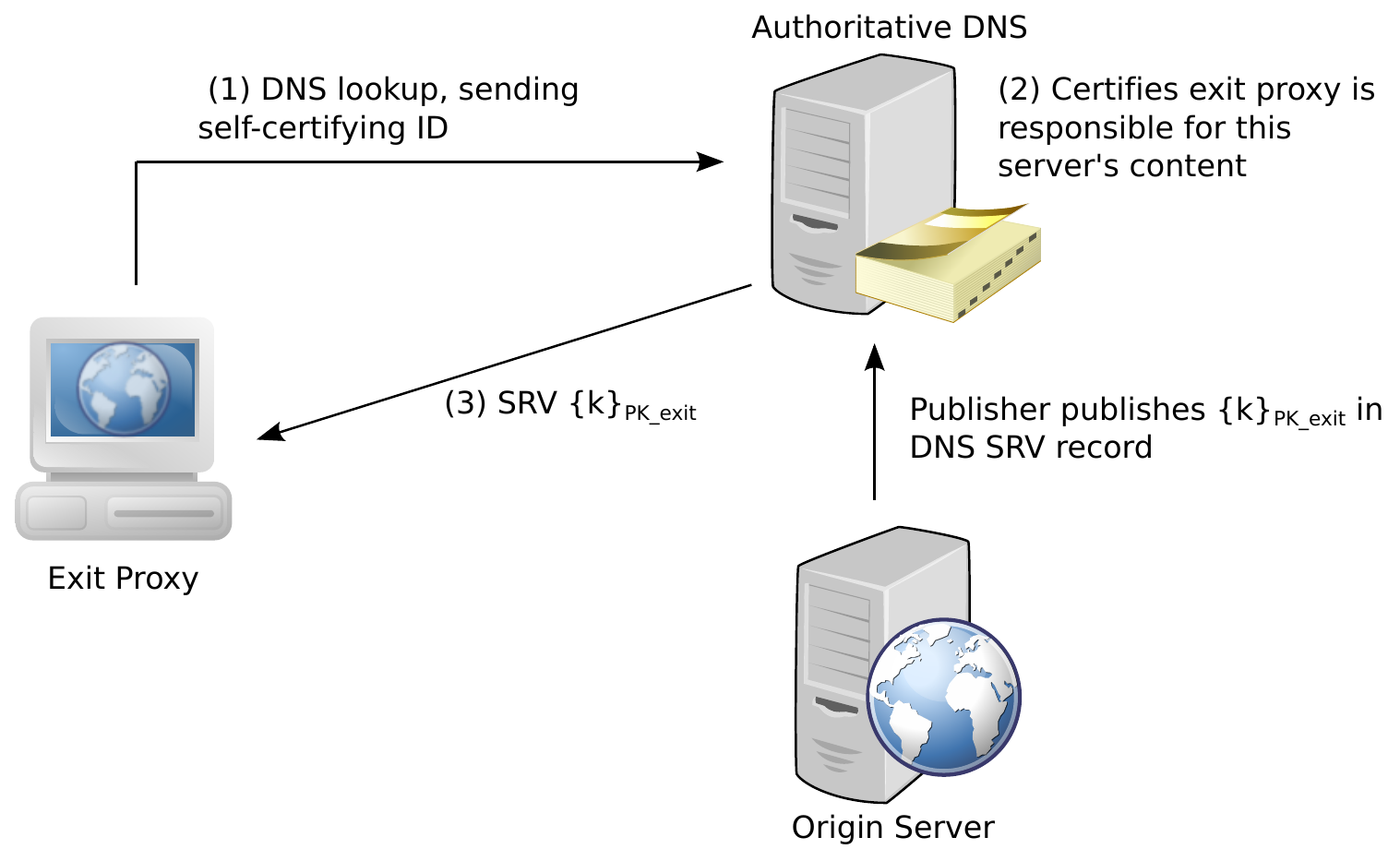}
\caption{How an origin server certifies an exit proxy and distributes its shared
key to an exit proxy.  In step~(1), 
the exit proxy sends his self-certifying ID in the {\it Additional} section of the DNS message.  }
\label{fig:keys}
\end{figure}

\section{Implementation}
\label{sec:implementation}

We have implemented a prototype of \system{} to demonstrate its feasibility and 
evaluate its performance.  Our implementation allows a client to send a request 
for content through an exit proxy, which will fetch the corresponding 
encrypted content.  Figure \ref{fig:impl} shows our prototype; the solid line represents
how \system{} communicates between the components, and the dotted line represents how 
a traditional CDN would communicate in our prototype.  Here we will discuss each component---client proxy, exit proxy, 
and CDN---separately, and how they fit together.

\begin{figure}[t]
\centering
\includegraphics[width=.45\textwidth]{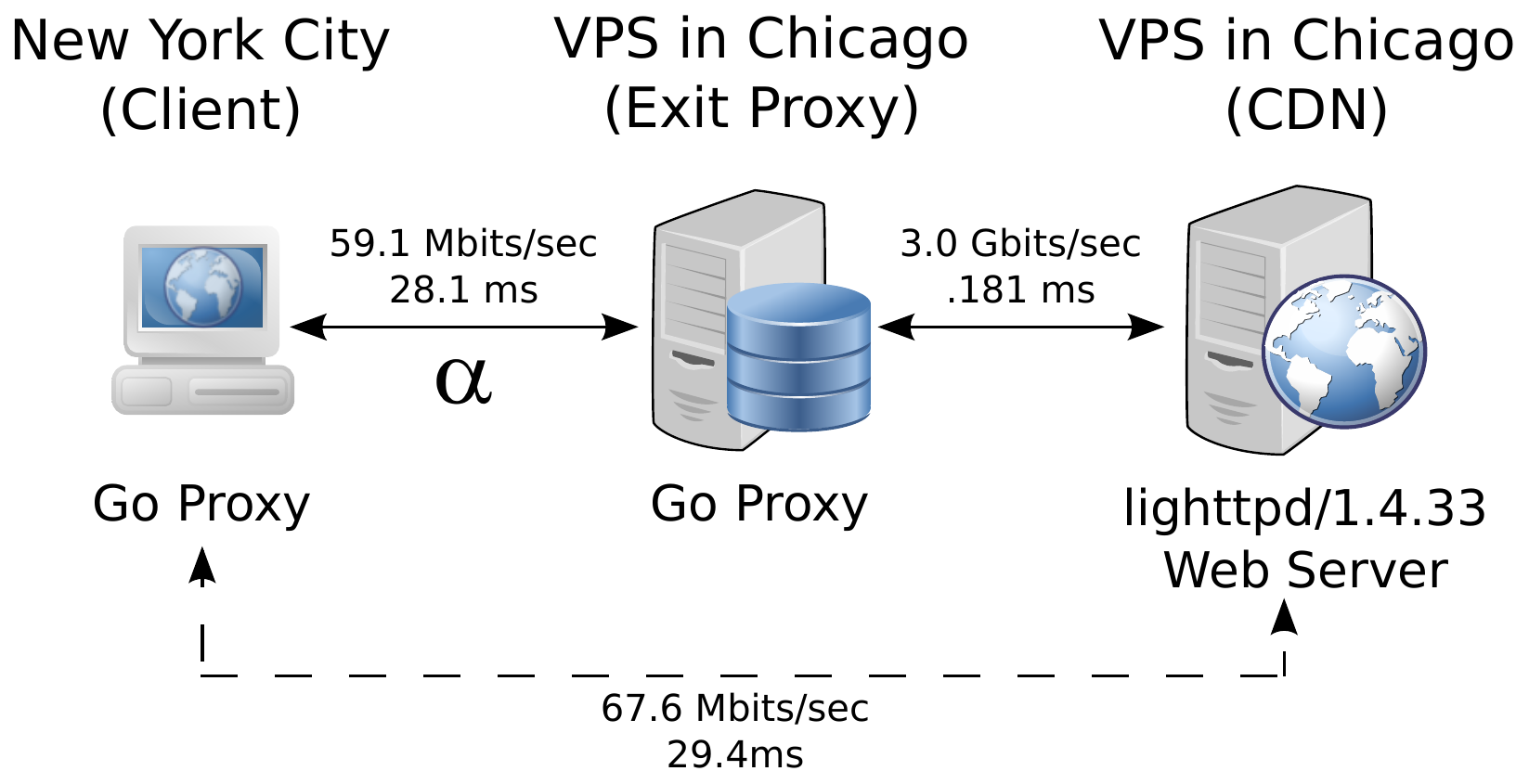}
\caption{The implementation of our \system{} prototype.  The solid line represents
how \system{} communicates between the components; the dotted line represents how 
a traditional CDN would communicate. $\alpha$ represents the latency between the client 
and the exit proxy; we simulate additional clients on this path by increasing $\alpha$.}
\label{fig:impl}
\end{figure}

\paragraph{CDN} As the design for \system{} requires encrypted content and identifiers
to be stored in the CDN, we cannot request content from real-world CDNs.  Additionally, 
we must evaluate the performance of \system{} in comparison to the same content, cache locations, etc., so 
we set up a data storage server.  This server is run on a Virtual Private Server
(VPS) located in
Chicago, USA.  To access content, we set up a web server on this VPS machine.  To generate 
plaintext web content, we used Surge~\cite{barford1998generating}, which allows us 
to generate a set of files that are representative of real-world web server file distributions.  
In \system{}, the files are encrypted with a shared key $k$ and the obfuscated file name is the 
HMAC$_{k}$(file name).  We use AES with 256-bit keys for the shared key and SHA-256
for the 
hash function.  Both the plaintext files and encrypted files are stored on this web server, and 
for the purposes of evaluating our prototype, act as a CDN in \system{}.

\paragraph{Exit Proxy} The exit proxy is the component that queries the CDN for
encrypted
content on behalf of a client.  We have implemented a web proxy in Go; this proxy
runs on
a different VPS machine in Chicago, USA.  In addition to proxying web requests, the exit 
proxy also provides cryptographic functionality.  When receiving a request, it rewrites
the URL in the request to be the HMAC$_{k}$(URL), and it parses the headers to retrieve a 
specific header, {\tt X-OCDN}, which contains the client's session key encrypted
under the exit
proxy's public key.  Our implementation uses 2048-bit RSA for asymmetric encryption.  After 
decrypting the session key, it stores it in memory for use on the response.  When 
it receives a response from the CDN, it decrypts the content with the shared key $k$, and 
subsequently encrypts it with the session key (both using AES 256-bit encryption).  The 
exit then forwards the response onto the client proxy.

\paragraph{Client Proxy} The client proxy acts on behalf of the client that is
requesting
content.  This proxy uses the same implementation as the exit proxy, but provides 
different cryptographic functions on the requests and responses.  When a client makes 
a request, the client proxy generates a session key  (AES 256-bit) and looks up the correct exit proxy's 
public key.  The client proxy then adds a header to the request, 
where X-OCDN is the key, the encrypted session key is the value.  The client then forwards this on to the 
exit proxy.  When the client receives a response from the exit proxy, it must decrypt the content 
with the session key it originally generated.   

\section{Security Analysis}
\label{sec:sec}
We analyze and discuss how \system{} addresses different attacks.  Table \ref{tab:sec_table} 
shows what security and privacy features \system{} provides in comparison to other related 
systems.

\begin{table}[t!]
\centering
\begin{tabular}{| l | c | c | c |} 
\hline
 {} & Preserves  & Preserves   & Protects \\ 
 {} & Integrity & Confidentiality & Client\\
 {} & at CDN & at CDN & Identity \\
\hline
 Stickler~\cite{levy2015stickler} & \checkmark & {} & {}\\ 
 R \& C~\cite{michalakis2007ensuring} & \checkmark & {} & {}\\
 Tor~\cite{dingledine2004tor} & {} & {} & \checkmark \\
 {\bf OCDN} & {} & {\bf \checkmark} & {\bf \checkmark} \\
\hline
\end{tabular}
\caption{The security and privacy features offered by related systems.  To our knowledge, 
\system{} is the first to address confidentiality at the CDN.}
\label{tab:sec_table}
\end{table}

\paragraph{Popularity Attacks}  An attacker that has requested or otherwise 
gained access to CDN cache logs can learn information about how often 
content was requested.  Because not all content is requested uniformly, the 
attacker could potentially correlate the most commonly requested content with 
very popular webpages.  While this does not allow the CDN to learn which 
clients are accessing the content, it can reveal information about what content 
is stored on the CDN cache nodes.  \system{} handles this type of attacker by making 
the distribution of content requests appear uniform.  The content publisher (of popular 
content) generates multiple encodings of their content and URLs, and encrypts each one 
with the shared key $k$, such that they have multiple, different-looking 
copies of their content.  All of the content copies are pushed to the CDN and the key is 
shared with the exit proxy.\footnote{This also provides load balancing for exit proxies 
that hold the shared key $k$ for the popular webpage because it distributes the load
across multiple exit proxies (where each of these exit proxies are responsible for 
one of the encodings).}  Now, the popular content does not appear as popular, 
and it makes it difficult for an attacker to infer the popularity of the content.

\paragraph{Chosen Plaintext Attacks} An attacker could attempt to
determine whether a particular URL was being accessed by sending requests
through specific \system{} proxies and requesting access to the CDN cache logs, 
which contain the corresponding obfuscated
requests and responses. Blinding the clients' requests
with a random nonce that is added by the proxy should prevent against this
attack. We also believe that such an attack reflects a stronger attack: from a
law enforcement perspective, receiving a subpoena for {\em existing} logs and
data may present a lower legal barrier than compelling a CDN to attack a
system.

\paragraph{Spoofed Content Updates} Because the CDN cache
nodes do not know either the content that they are hosting or the URLs
corresponding to the content, an attacker could masquerade as an origin server
and could potentially push bogus content for a URL to a cache node. There are
a number of defenses against this possible attack. This simplest solution is
for CDN cache nodes to authenticate origin servers and only accept updates
from trusted origins; this approach is plausible, since many origin servers already
have a corresponding public key certificate through the web PKI hierarchy.  An additional
defense is to make it difficult for to discover which obfuscated URLs correspond
to which content that an attacker wishes to spoof; this is achievable by design.
A third defense would be to only accept updates for content from the same origin
server that populated the cache with the original content.

\paragraph{Flashcrowds}  A flashcrowd is large spike in traffic to a specific web
page. An attacker
could see that some content on the CDN has just seen a surge in traffic and correlate that with 
other information (for example, major world events).  This leaks information about what content the 
CDN is caching.  Fortunately, the design of \system{} can defend against this type of inference attack.  
The exit proxy can cache content in the time of a flashcrowd, such that the CDN (and therefore the attacker) 
does not see the surge in traffic.\footnote{This raises billing issues because the CDN can’t charge as much if edge servers don’t see as many requests for the origin; fortunately, RFC 2227 describes a solution for this~\cite{rfc2227}.}  

\section{Performance Analysis}
\label{sec:performance}
To evaluate how much overhead is caused by \system{} we measure the performance 
of \system{}.  In addition to understanding the latency and overhead produced by the 
system, we also discuss the scalability of the design and show how \system{} scales 
well with an increasing number of clients.

\subsection{\system{} Overhead}
For measuring performance characteristics of \system{}, we use the implementation 
described in Section \ref{sec:implementation}.  Figure \ref{fig:impl} shows 
how our measurements reflect \system{} (solid line) and a traditional CDN (dotted 
line).  

Figure \ref{fig:ttfb} shows the Time to First Byte (TTFB) for both \system{} and 
without \system{}.  We can see the the TTFB using \system{} grows linearly with 
file size, whereas without \system{} TTFB remains fairly constant.  Interestingly, 
we can see that there are some fixed time operations that \system{} performs, which 
is visible by looking at the smaller file sizes.

\begin{figure}[t!]
\centering
\includegraphics[width=.5\textwidth]{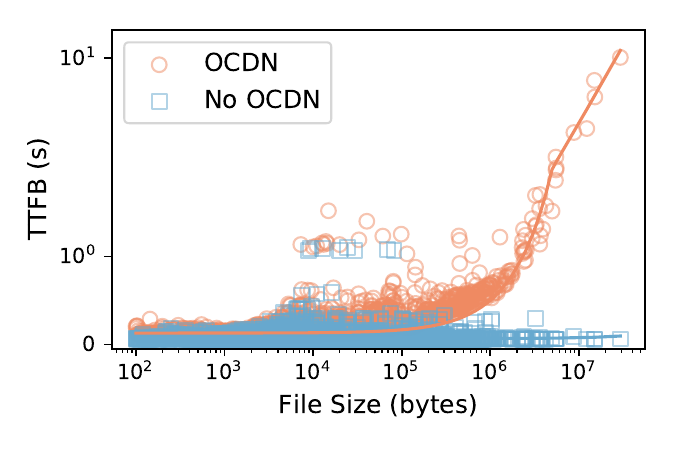}
\caption{Time to First Byte measurements with and without \system{}.}
\label{fig:ttfb}
\end{figure}

In addition to measuring TTFB, we measured the time it took to complete a request (with and 
without \system{}); the results are shown in Figure \ref{fig:completion}.  Again, completion time 
grows linearly with file size, but for both \system{} and without \system{}; while both follow the 
same pattern, the time to complete requests is, as expected, longer using \system{} as it performs 
many cryptographic operations and proxies traffic between the client and the CDN.  

\begin{figure}[t!]
\centering
\includegraphics[width=.5\textwidth]{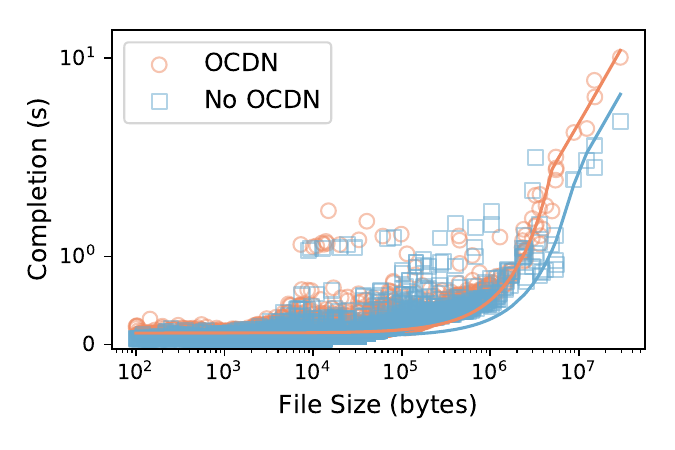}
\caption{Time to complete a request with and without \system{}.}
\label{fig:completion}
\end{figure}

As described in Section \ref{sec:implementation}, our prototype included only a single client, but 
our design allows for a client to proxy her request through additional clients.  To simulate this, we 
add latency between the client and the exit proxy, and measure both the TTFB and time to complete a request 
when there are different values of latency, which represent different numbers of clients on the path between the 
original client and the exit proxy.  Figure \ref{fig:latency} shows the results for three different file sizes. The 
bottom portion of each bar in the graph shows the TTFB, and the top portion shows the additional time needed 
to complete the request.  As expected, the TTFB grows much slower as file size and latency increase; completion time 
grows more quickly than TTFB as the file size and latency increase.   

\begin{figure}[t!]
\centering
\includegraphics[width=.5\textwidth]{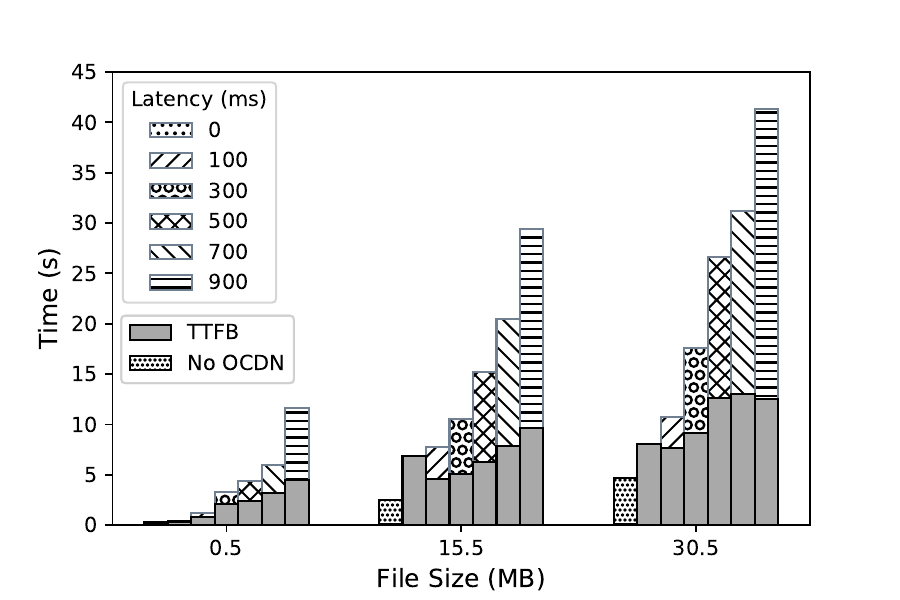}
\caption{Time to First Byte and time to complete a request with varying the file size and latency; this latency 
correspondes to $\alpha$ in Figure \ref{fig:impl}.}
\label{fig:latency}
\end{figure}

Finally, we measure the performance overhead of the individual operations used in
\system{}; figure \ref{fig:overhead2} shows
the overhead of different components of the system for three different file sizes.  We can see that some of the fixed cost/time 
operations include the client locally looking up the correct exit proxy to use for a given URL and the exit proxy generating the 
HMAC$_{k}$(URL).  The operations that have the most overhead and continue to grow with the size of the file are the exit proxy decrypting 
the response with the shared key $k$, the exit proxy encrypting the response with the session key $k_{session}$, and the client 
decrypting the response with the session key $k_{session}$.

\begin{figure}[t!]
\centering
\includegraphics[width=.5\textwidth]{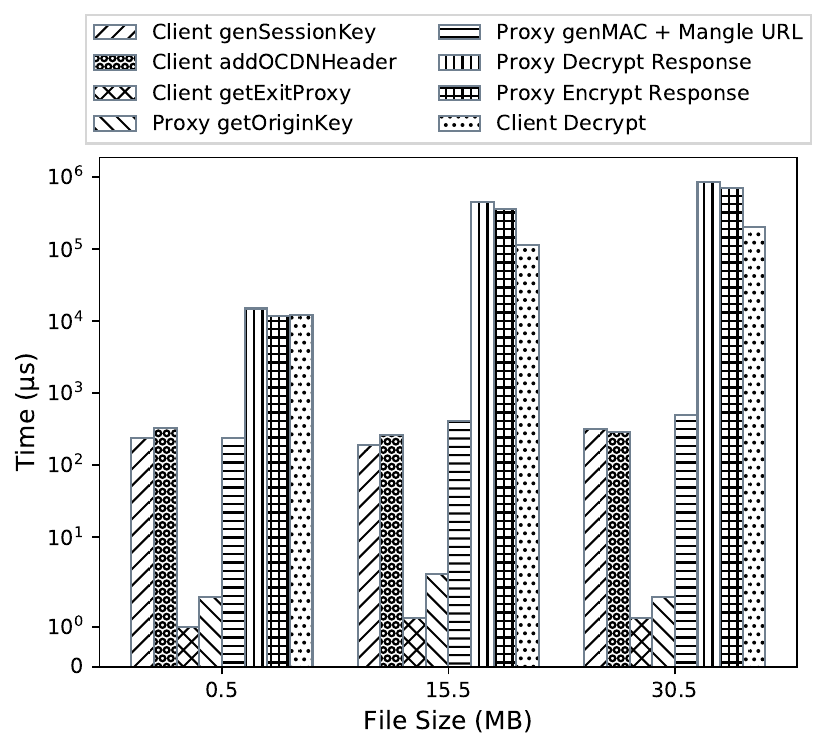}
\caption{Overhead of different operations performed by \system{}.}
\label{fig:overhead2}
\end{figure}

\subsection{Scalability}
For evaluating performance, we are also concerned with how well \system{} will scale with more users 
and more URLs.  In particular, we need to reason about how much load is put on the exit proxies as the 
system grows; clients do not bare much load in the system as they simply proxy requests and the CDN is designed 
to handle high numbers of requests, therefore, we limit our scalability analysis to the exit proxies.  

As previously mentioned, we balance load among the proxies by using consistent hashing to assign URLs to 
exit proxies.  \system{} can additionally distribute load by replicating exit proxies, meaning that two exit proxies can 
have the same distributed hash table of shared keys.  This way, both exit proxies can accept requests from clients for 
the URLs they are responsible for.  Also worth noting is that the exit proxy is only recieving requests for the content 
corresponding to the shared keys it contains.  Therefore, as the number of clients grow, the exit proxy is still only responsible 
for its set of shared keys and subsequent URLs.  And as the number of URLs increase, the additional load per proxy is 
still low because of the load balancing properties of consistent hashing.  We also discuss in Section \ref{sec:discussion} how 
clients can set up exit proxies; this will further decrease the load per exit proxy because each exit proxy will be responsible 
for a smaller number of shared keys/URLs.

\section{Discussion}
\label{sec:discussion}

In this section, we discuss the various technical, political, and legal limitations
of \system{}, as well as possible avenues for future work. 

\paragraph{\system{} limitations} CDNs become slightly limited in terms of the 
possible performance optimizations when following \system{}'s design.  For example, 
many CDNs perform HTTPS re-writes on content that they cache, but this can only be 
done if the CDN has access to the decrypted content.  Similarly, the CDN needs the 
decrypted content to perform minimizations on HTML, CSS, and Javascript files.  While 
this likely increases performance in traditional CDNs, it does not provide the greatest 
increase in performance; content caching around the world is the greatest benefit to 
performance, which \system{} preserves.

\paragraph{CDNs operated by content hosts} The design of \system{}
assumes that the entities operating the proxies and delivering content are
distinct from original content provider. In many cases, however---particularly
for large content providers such as Netflix, Facebook, and Google---the
content provider operates their own CDN cache nodes; in these cases, \system{} will
not be able to obfuscate the content from the CDN operator, since the content host
and the CDN are the same party.  Similarly, because the CDN operator is the same
entity as the original server, it also knows the keys that are shared with the clients.
As a result, the CDN cache nodes could also discover the keys and identify both
the content, as well as which clients are requesting which content.

\paragraph{Exit proxies run by volunteers} In the description of \system{} in Sections 
\ref{sec:design} and \ref{sec:protocol}, we assume we are running the exit proxies, but 
the design of the system also allows for clients to run exit proxies.  Exit proxies are not 
trusted with client identities and information, which allows for volunteers to run exit proxies.  
The addition of an exit proxy follows the protocol in consistent hashing for when a new node 
joins; some keys would be transferred to the new exit proxy, and clients' mapping of 
exit proxies will be updated.  This allows for the load to be split among more proxies and 
increases the geographic diversity of the exit proxies.


\paragraph{Legal questions and political pushback} Recent cases surrounding
the Stored Communications Act in the United States raise some questions over
whether a system like \system{} might face legal challenges from law
enforcement agencies. To protect user data against these types of challenges,
Microsoft has already taken steps such as moving user data to data centers in
Germany that are operated by entities outside the United States, such as
T-Systems. It remains to be seen, of course, whether \system{} would face similar
hurdles, but similar systems in the past have faced scrutiny and pushback from law
enforcement.

\section{Related Work}
\label{sec:related}

To our knowledge, there has been no prior work on preventing surveillance at CDNs, but there 
has been relevant research on securing CDNs, finding security vulnerabilities in CDNs, and 
conducting different types of measurements on CDNs.

\paragraph{Securing CDNs} Most prior work on securing CDNs has focused on providing content 
integrity at the CDN as opposed to content confidentiality (and unlinkability).  In 2005, 
Lesniewski-Laas and Kaashoek use SSL-splitting --- a technique 
where the proxy simulates an SSL connection with the client by using authentication records from 
the server with data records from the cache (in the proxy) --- to maintain the 
integrity of content being served by a proxy~\cite{lesniewski2005ssl}.  While this work does not 
explicitly apply SSL-splitting to CDNs, it is a technique that could be used for content 
distribution.  Michalakis et al., present a system for ensuring content integrity for untrusted 
peer-to-peer content delivery networks~\cite{michalakis2007ensuring}.  This system, Repeat and 
Compare, use attestation records and a number of peers act as verifiers.  More recently, Levy et al., 
introduced Stickler, which is a system that allows content publishers to guarantee the end-to-end 
authenticity of their content to users~\cite{levy2015stickler}.  Stickler includes content publishers 
signing their content, and users verifying the signature without having to modify the browser.  Unfortunately, 
systems like Stickler do not protect against an adversary that wishes to learn information about content, clients, 
or publishers; \system{} is complementary to Stickler.

There has been prior work in securing CDNs against DDoS attacks; Gilad 
et al. introduce a DDoS defense called CDN-on-Demand~\cite{gilad2016cdn}.  In this work they 
provide a complement to CDNs, as some smaller organizations cannot afford the use of CDNs and 
therefore do not receive the DDoS protections provided by them.  CDN-on-Demand is a software 
defense that relies on managing flexible cloud resources as opposed to using a CDN provider's 
service.\\

\paragraph{Security Issues in CDNs} More prevalent in the literature than defense are attacks on CDNs.  Recent work 
has studied how HTTPS and CDNs work together (as both have been studied extensively separately).  Liang et al., studied 
20 CDN providers and found that there are many problems with HTTPS practice in CDNs~\cite{liang2014https}.  Some of these 
problems include: invalid certificates, private key sharing, neglected revocation of stale certificates, and 
insecure back-end communications; the authors point out that some of these problems are fundamental issues due to 
the man-in-the-middle characteristic of CDNs.  Similarly, Zolfaghari and Houmansadr found problems with HTTPS usage by 
CDNBrowsing, a system that relies on CDNs for censorship circumvention~\cite{zolfaghari2016practical}.  They found that HTTPS 
leaks the identity of the content being accessed, which defeats the purpose of a censorship circumvention tool. 

Research has also covered other attacks on CDNs, such as flash crowds and denial of service attacks; Jung et al., show 
that some CDNs might not actually provide much defense against flash events (and they differentiate flash events from denial 
of service events)~\cite{jung2002flash}. Su and Kuzmanovic show that some CDNs are more susceptible to intential service 
degradation, despite being known for being resilient to network outages and denial of service attacks~\cite{su2008thinning}. 
Additionally, researchers implemented an attack that can affect popular CDNs, such as Akamai and Limelight; this attack 
defeats CDNs' denial of service protection and actually utilizes the CDN to amplify the attack~\cite{triukose2009content}.  In the 
past year, researchers have found forwarding loop attacks that are possible in CDNs, which cause requests to be served repeatedly, which 
subsequently consumes resources, decreases availability, and could potentially lead to a denial of service attack~\cite{chen2016forwarding}.

Recently, researchers have studied the privacy implications of peer-assisted CDNs; peer-assisted CDNs allow clients to cache and distribute 
content on behalf of a website.  It is starting to be supported by CDNs, such as Akamai, but the design of the paradigm
makes clients susceptible to privacy attacks; one client can infur the cross site browsing patterns of another client~\cite{jia2016anonymity}.\\

\paragraph{Measuring and Mapping CDNs} As CDNs have increased in popularity, and are predicted to grow even more~\cite{predict}, much research has 
studied the deployment of CDNs.  Huang et al., have mapped the locations of servers, and evaluated the server availability for two CDNs: 
Akamai and Limelight~\cite{huang2008measuring}.  More recently, Calder et al., mapped Google's infrastructure; this included 
developing techniques for mapping, enumerating the IP addresses of servers, and identifying associations between clients and clusters of 
servers~\cite{calder2013mapping}. Scott et al., develop a clustering technique to identify the IP footprints of CDN deployments; this analysis
 also analyzes network-level interference to aid in the identification of CDN deployments~\cite{scott2016satellite}.  In 2017, researchers 
conducted an empirical study of CDN deployment in China; they found that it is significantly different than in other parts of the world 
due to their unique economic, technical, and regulatory factors~\cite{xue2017cdns}. 

Other measurement studies on CDNs have focused on characterizing and quantifying flash crowds on CDNs~\cite{wendell2011going}, inferring 
and using network measurements performed by a large CDN to identify quality Internet paths~\cite{su2009drafting}, and measuring object size distributions and 
request characteristics to optimize caching policies~\cite{berger2017adaptsize}.

\balance\section{Conclusion}
\label{sec:conclusion}

As more content is distributed via CDNs, CDNs are increasingly becoming the
targets of data requests and liability cases.  We discuss why CDNs are
powerful in terms of the information they know  and can gather, such as a
client's cross site browsing patterns.  In response to  traditional CDNs'
capabilities, we design \system{}, which provides oblivious content
distribution.  \system{} obfuscates data such that the CDN can operate without
having knowledge of  what content they are caching.  This system not only
provides protections to CDNs, but also preserves client privacy by ensuring
that the CDN and the origin server never learn the identity of clients that
make requests for content. \system{} is robust against a strong adversary who
has access to request logs at the CDN and can also join the system as a
client, publisher, or CDN. Our evaluation shows that \system{} imposes some performance
overhead due to the cryptographic operations that allow it to obliviously cache
content, but that this overhead is acceptable, particularly for the sizes of files
that make up common web workloads.
\label{lastpage}

\end{sloppypar}


\small
\balance\bibliography{paper}
\bibliographystyle{abbrv}
}{
}
\pagebreak

\end{document}